# The random model for simulation of the growth of small populations[*]

Kazimierz Pater

*Institute of Physics, Wrocław University of Technology, Wyb. Wyspiańskiego 27,
50-370 Wrocław, POLAND*
pater@if.pwr.wroc.pl

**Abstract**

The probability of the survival of the population of individuals of both sexes of given mature age, procreation rate and structure stability has been searched in the numerical experiment. The populations with long period of reproduction and the high rate of procreation and without social mobility have the most chance to survive. The populations with the late mature age and high mobility dies out. The fertility rate of simple reconstruction of generations obtained in the model (2.8) is close to the value for a human being (2.1).

## 1  Introduction

In the very beginning, I would like to make it clear that the subject of my lecture is mainly the model. Thus, I am not going to present any formulas or theoretical considerations. I have read in a well-known book written by Stephen Hawking [1] that the presence of only one formula in a book decreases its sale by 50%. Perhaps it is the reason why very few people read physics handbooks. In fact, the main reason is the lecture of professor Sznajd entitled "Power laws – the Saint Grall of the complex systems", which I have heard. Professor Sznajd (known as a co-author of a well-known socio-physical model [2]) claims that it is relatively easy to introduce for example, a law describing relation between the size of a town and a selected order parameter. However, nothing results from such a law - we obtain a pure phenomenological formula. Yet, science begins at the point where we are trying to answer the question: "Why?" In this meaning this paper is not a scientific work. It is only a presentation of a certain experiment which deals with a huge number of random choices, possible to execute with PC. In fact, it can be done even by using only a dice and having a lot of time.

Here, I would like to present the genesis of the model, the way its premises can be proved and, of course, its results, which will be treated rather roughly. It is a demographic model and it is "unintelligent," which means that individuals are subjected entirely to random laws and they are not governed by any personal motivations. Besides, it is rather a "crude" model and its population is artificial, although I attempted to include in it possibly real parameters, characteristic of small groups of individuals of both sexes.

---

[*] The lecture held on the conference "Sociophysics – Bielefeld 6-9 June 2002"

## 2 Model

To examine a population growth, the Penna model [3] with its various modifications reviewed by Sauffer [4] is widely used. It is a genetic model in which an important fact is the transmission of the information about the quality of genotype of an individual to offspring, which determines the chance for its survival and the transfer of good genes to the next generations. An unquestionable success of the Penna model is its agreement (Fig.1) with the Gompertz law describing the dependence of mortality on age. Both, the Penna model and the Gompertz experimental law known from the XIXth century, show the exponential growth of mortality vs. age.

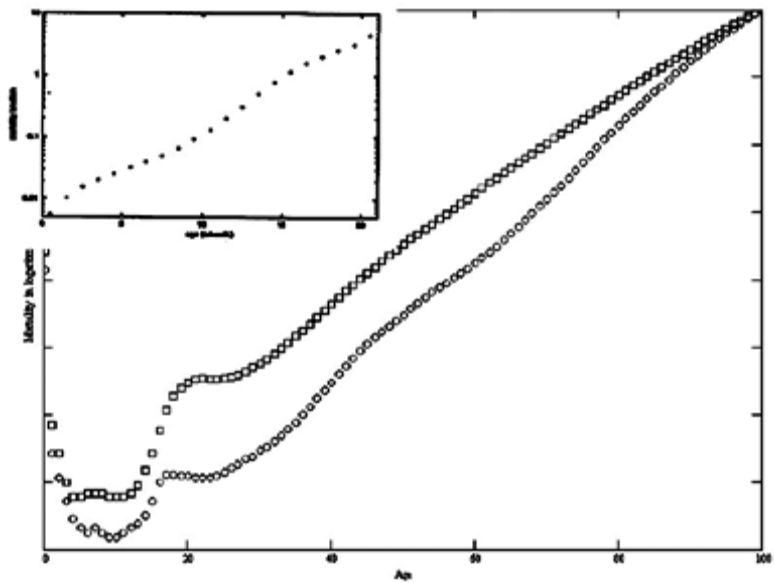

Figure 1: Semilogarithmic plot of mortality functions of men (sqares) and women
(circles) in Poland 2000 from Statistical Central Bureau of Poland
(http://www.stat.gov.pl).
In the left upper part, simulated mortality function in asexual Penna model [4].

However, it does not work during the initial part of a procreation period. Nevertheless, this property of the real populations, as well as the asexuality of the Penna model along with its later modifications forced me to search an alternative suggestion. I decided to design a model for searching the multigenerational growth of a population and I focused on the reproduction abilities of the population, treating the problem of dying as a minor one. Considering the problem from this point of view, it can be assumed that dinosaurs died out not because of a world catastrophe but because they ceased to reproduce.

I have supposed that the model (Fig.2) should enable the sexual reproduction. Therefore, it should ensure joining two individuals of different sex and production of offspring with a certain probability. It is also obvious that the model should ensure the natural growth of individuals and their death. Another important feature of population seems to be its internal structure: hierarchical, casual or random. I have noticed that the social structure of populations that I know contains gaps. For example, a given individual should appear singly when he/she is not attractive enough as a partner to procreation or if his/her partner died. Social structures are multidimensional, though in the models known to me, usually one or two-dimensional (geographic) set of relations between individuals is accepted. In my paper, the one-dimensional model is proposed. In such a model it is easy to introduce a hierarchy and the couple matching of close neighbors is possible. The least information about individual is the information about his age. It allows to assume the minimum and maximum procreation ages. Moreover, it is assumed that the set of numbers determining the ages of the individuals is the least information about given population needed to examine its durability. If we place some zeros in the set, it will make possible to create gaps in the social structure. The vector made of such numbers is taken as the representation of population at a given stage of its evolution.

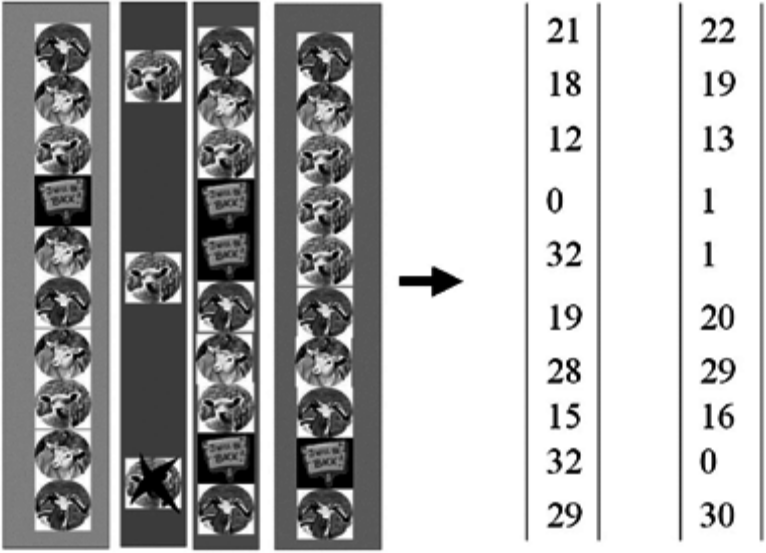

Figure 2: Numerical representation of the evolution of artifical groups of individuals of both sexes.

In the proposed model, the natural way of determining the sex of an individual is the parity of the number which represents it. It does not mean that the sex of all individuals changes every year. The kind of sex is necessary only at the moment of couple matching and it is important for the sexes to be different, in order to ensure parity control. Anyway, the condition of procreation including different sexes may became unnecessarily in the future. We all must die, which is a sad truth with no exceptions so far. As it has been already mentioned, the main field of my interest is the procreation itself so I treat the question of death of any individual rather radically. Everyone dies when the procreation age, common for all individuals, ends. Of course, it is relatively simple to include in the model the Verhulst's factor or a death resulting from a random accident. Yet, having in mind the mentioned departure from the Gompertz's law at the phase of procreation, I assumed the above solution, which can be also interpreted less strictly, as a way out of the procreation period. Another, separate and even more important problem is the birth of a new member of a group of individuals and his location in the group's structure. I assumed that the newborns are located first in the empty spaces of the structure or, in the case of the lack of gaps in it, increase the number of the population and form the cluster of peers added at the end of the structure. The creation of such clusters is, as we will see later, very important for population permanence. Such a way of the population increase ensures its compactness and, from my point of view, allows to reproduce its structure. The group structure may remain intact i.e. "as grow", and can be randomly changed, as well as forced. All three possibilities were used in the model and it came out that they had really significant influence on the persistence of the population. The history of any given population is not of my interest. I only assume that a population will either grow or die. My aim is to examine the chances for survival of a population with such attributes like: the beginning of procreation, the probability of producing offspring by a pair of parents of different sex, and the probability of changing location in the group structure. I assumed that a population survives if its size is ten times bigger than the size of the initial one. I have also investigated the influence of hierarchy existing in the group on the chances of its survival. This is the reason why I have introduced in B version of the model a forced age hierarchy ordering structure of the group - from the oldest to the youngest individual.

The program executing the assumptions mentioned above has been written in quite an exotic language and I realize that more effective procedures can be created. The program is compact, although the simulation of one population set with a common period of procreation took about 24 hours for ordinary PC. The program drew 100 populations of fixed parameters and examined the number of populations which survived. In such a simple way the percentage of chances of the survival of a population was calculated. It was also possible to calculate time (number of time steps) in which population reached permanent growth (i.e. survived) or disappeared. The program ensured the population rate independent of the number of its individuals.

## 3 Results and discussion

The results of the simulation are in accordance with expectations (Fig.3). The biggest chance for survival have the populations with the early age of procreation beginning, high probability of offspring production, and showing no changes inside the group, i.e. "as grow". The populations with opposite features, particularly those with the maximum social movements have no chances for survival, regardless of the forced age hierarchy (model B) or without any hierarchy (model A).

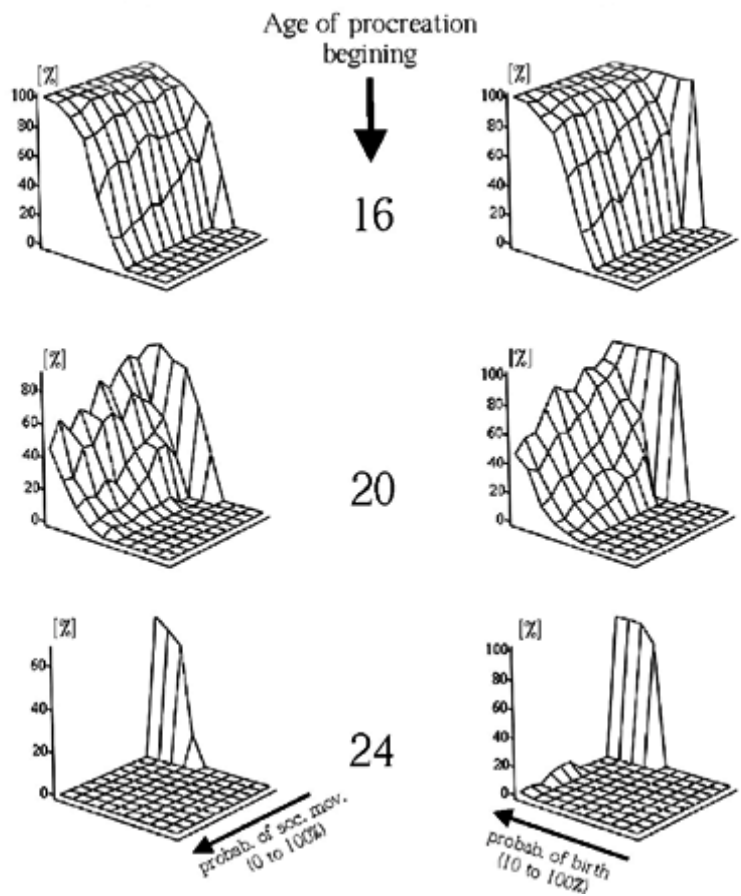

Figure 3: Survival probability of model without forced hierarchy (left) and with hierarchy (right).

The only one clear effect of hierarchy introduction is the distinction of transition between the populations without chances for survival and those that survived. In such cases physicists use a term "phase transition". Such a transition is clearly observed only for "as grow" populations and it relates to forming of "clusters of one-generation individuals" in the group structure. In such clusters the chances to find a partner for procreation are bigger than in the disordered structure. Including hierarchization intensifies the possibility of faster population growth (or death) and, what is interesting, it does not lower the threshold of phase transition. It turns out that even the weakest mobility inside the group (10 % of population) destroys the clusters (also in the B model). In my opinion the term "phase transition" could be accepted for the "as grow" populations because, in this case, we can see the transition from the structure consisting of the individuals randomly distributed to the structure consisting of the clusters of peers i.e. "disorder – order". In order to define the parameters of that transition, the lifetimes (or the times of achieving the phase of permanent growth) of all "as grow" populations of A and B models were analyzed.

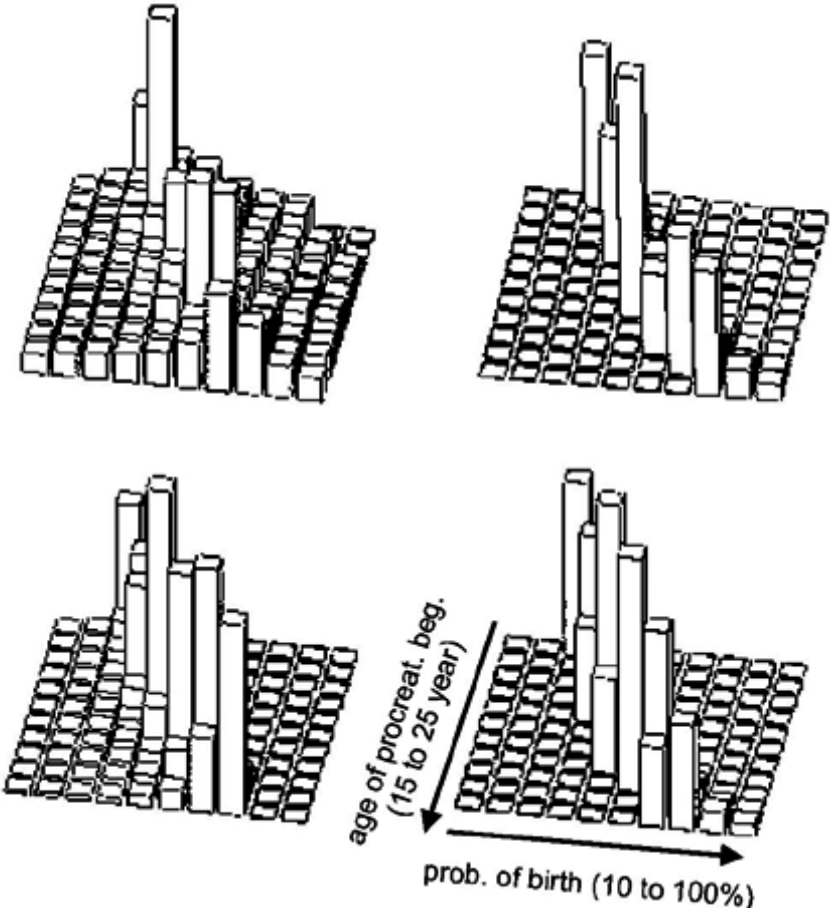

Figure 4: Lifetime (right) or stable growth time achievement (left) of different kind of populations – model A (above), model B (below).

As it is presented in Fig.4, for those times the clear maximum is observed for the probability of birth which depends on the age of procreation start. From the data presented in this figure the threshold of population rate can be estimated at 2.2 ± 0.2% for the both models A and B. This number is at least one order of power higher than a standard population rate for people but the fertility rate of simple reconstruction of generations calculated by using that value is 2.8 ± 0.2. This value is very close to the value for a human being 2.1 [5] but I suppose that it is a sheer coincidence.

The results obtained should be treated rather carefully. After all, it is only a simulation, something like a "gedanke" experiment which significance is rather little. It is a kind of game and we still follow Einstein's idea that "Good does not play dice" and life processes are not random.

Answering the question asked by professor Stauffer: " Why is the mortality difference between that for men and women in Poland so big?" I said that I did not know exactly but the probable reason is that men work harder, which was accepted by the audience (men mainly).